\documentclass[submission,copyright,creativecommons]{eptcs}
\providecommand{\event}{TTC 2013}
\usepackage[english]{babel}
\usepackage[latin1]{inputenc}	
\usepackage[T1]{fontenc}
\usepackage{graphicx}
\usepackage{url}
\usepackage{llncsdoc}
\usepackage{tabularx}
\usepackage{amsmath}
\usepackage{amssymb}
\usepackage{enumerate}
\usepackage{scrtime}
\usepackage{listings}
\usepackage{subfigure}
\usepackage{hyperref}
\usepackage{float}

\hypersetup{
 pdfauthor={Georg Hinkel},
 pdftitle={NMF Solution for Flowgraph case},
 pdfsubject={TTC2013},
 pdfkeywords={MDSD,MDE}
}

\title{An \textsc{NMF} solution for the \textit{Flow Graphs} case study at the TTC 2013}
\author{Georg Hinkel
\institute{Karlsruhe Institute of Technology \\ Karlsruhe, Germany}
\email{georg.hinkel@student.kit.edu}
\and
Thomas Goldschmidt
\institute{ABB Corporate Research \\ Ladenburg, Germany}
\email{thomas.goldschmidt@de.abb.com}
\and
Lucia Happe
\institute{Karlsruhe Institute of Technology \\ Karlsruhe, Germany}
\email{lucia.kapova@kit.edu}
}
\def\titlerunning{An \textsc{NMF} solution for the \textit{Flow Graphs} case study at the TTC 2013}
\def\authorrunning{G. Hinkel, T. Goldschmidt \& L. Happe}

\begin{document}
\selectlanguage{english}

\maketitle


\begin{abstract}
Software systems are getting more and more complex. Model-driven engineering (MDE) offers ways to handle such increased complexity by lifting development to a higher level of abstraction. A key part in MDE are transformations that transform any given model into another. These transformations are used to generate all kinds of software artifacts from models. However, there is little consensus about the transformation tools. Thus, the Transformation Tool Contest (TTC) 2013 aims to compare different transformation engines. This is achieved through three different cases that have to be tackled. One of these cases is the Flowgraphs case. A solution has to transform a Java code model into a simplified version and has to derive control and data flow. This paper presents the solution for this case using NMF Transformations as transformation engine.
\end{abstract}


\section{Introduction}
\label{ch:Introduction}

The challenge of the Flowgraphs Case \cite{ttcFlowGraphs} is to derive a control flow graph and data flow graph from a JaMoPP \cite{heidenreich2009jamopp} model representing Java code. The case is divided into four subtasks. The first task deals with the creation of an initial flow graph model out of the JaMoPP model representing the Java program. The second task is to derive the control flow within this flow graph model. The third task deals with the issue to derive the data flow out of the prior. Task four finally demands a tool to validate the output of the previous transformations. The transformation tasks are tackled with \textsc{NMF}\footnote{\url{http://nmf.codeplex.com/}}, an open source project to support model-driven engineering on the .NET platform. The solution is available on SHARE\footnote{\url{http://is.ieis.tue.nl/staff/pvgorp/share/?page=ConfigureNewSession&vdi=XP-TUe_TTC13::NMF_TTC13::NMF_updated_NMF_LiveContest.vdi}}.

\section{.NET Modelling Framework (\textsc{NMF})}
\label{ch:NMF}

The .NET Modelling Framework is an open source project that provides support for model-driven software development on the .NET platform. An essential part is the model transformation engine, \textsc{NMF Transformations}, which allows to write rule-based transformations in arbitrary .NET languages using an internal DSL \cite{fowler2010domain}. The reason to implement the transformation language as internal DSL is mainly that transformation languages ought to be Turing complete \cite{sendall2003model} and thus, many advantages of external DSLs attenuate. An internal DSL, however, can make use of features of its host language. Developers used to this language feel familiar with the DSL.

\textsc{NMF Transformations} makes it possible to specify model transformations directly in C\#. For this purpose, \textsc{NMF Transformations} has a simple abstract syntax but hides the complexity in the attributes of the metaclasses which are representing functions. These functions can be specified with general purpose code that contain code as sophisticated as required. Although the transformation language might seem quite verbose, especially when compared with external model transformation languages, C\# has been chosen as host language to make it easier to write and thus maintain these transformations for C\# developers.

Currently, \textsc{NMF} does not contain a metamodeling foundation, e.g. based on MOF. Instead, \textsc{NMF Transformations} uses the concepts of the CLR (the virtual machine used on the .NET platform) to represent models and operates on plain objects (POCOs). Thus, we used an interop component to EMF, which generates classes from an Ecore metamodel. Furthermore, there exists a serializer component to load and store simple models (models that do not have references to other files).

Beside \textsc{NMF}, to the best of our knowledge, there is hardly any framework that supports model-driven engineering on the .NET platform. Microsoft offers a Visualization and Modeling SDK\footnote{\url{http://archive.msdn.microsoft.com/vsvmsdk}} for Visual Studio, which is a tool for graphical editors, and T4 as Text-To-Text-Transformation engine. However, although T4 is included in Visual for many years now, there is still hardly support for editing T4 templates. There is an add-in providing syntax-highlighting and code-completion, but still the support is much less than for writing normal e.g. C\# code. Furthermore, T4 has some restrictions like no inheritance is allowed within a T4-template. These restrictions and the lack of out-of-the-box tool support make model-driven development hard. \textsc{NMF Transformations} makes it possible to use the full tool support for C\# also for model transformations.

\section{Solution}
\label{ch:Solution}

All of the tasks have been tackled. The final solution consists of two separate C\# console projects. The first console application reads a JaMoPP file, transforms it to a Structure Graph (Task~1, see section \ref{ch:Solution:Task1}), derives the control flow (Task~2, see section \ref{ch:Solution:Task2}) and sets the data flow (Task~3.2, see section \ref{ch:Solution:Task3:Subtask3.2}). Each of the transformations operates in memory, only. After all transformations have been applied, the resulting data flow model is persisted using XMI into an output file specified by command line parameters. Furthermore, deriving the control flow graph and setting the data flow can be switched on or off using command line arguments. The second console application validates the links (see section \ref{ch:Solution:Task4}).

\subsection{Task 1/3.1: Structure Graph}
\label{ch:Solution:Task1}

A M2M-transformation in \textsc{NMF Transformations} is specified through transformation rules, which are represented by classes. These transformation rules may only be called once per input arguments in a transformation context. This context represents a transformation pass and provides trace functionality. Furthermore, transformation rules may define dependencies to other rules. These dependencies are necessary to set, in order to let \textsc{NMF} Transformations know which rules to call and to derive the inputs of these rules. \textsc{NMF Transformations} operates on plain CLR objects and therefore does not know the structure of the metamodel. Thus the structure of the domain model has to be reflected in the dependencies of the transformation rules. The only rule that is actually called by \textsc{NMF Transformations} automatically is the rule that matches the transformation request to transform a Java code model into a flow graph.

\begin{figure}[ht]
		\includegraphics[width=1.00\textwidth]{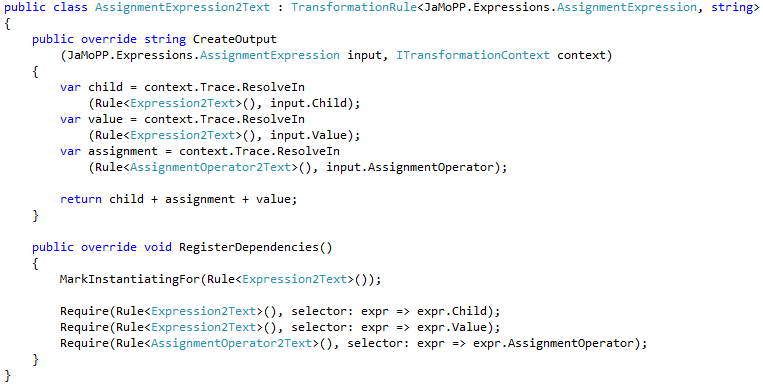}
	\caption{The transformation of assignment expressions}
	\label{fig:assignmentExpression2text}
\end{figure}

Due to space limitations, only an example of the transformation rules involved in the initialization can be shown here. Figure \ref{fig:assignmentExpression2text} shows the transformation of \textit{AssignmentExpression} elements. The code within \textsl{RegisterRequirements} shows how dependencies are specified using lambda expressions that define which subsequent elements to call. This correspondence can be queried later on. As the target model for expressions is plain strings, which are immutable in .NET, the transformation efforts need to be done within \textsl{CreateTransformationOutput}. 

\begin{figure}[ht]
	\centering
		\includegraphics[width=0.40\textwidth]{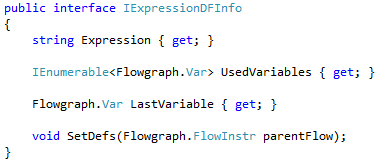}
	\caption{The interesting attributes for an Expression}
	\label{fig:IExpressionDFInfo}
\end{figure}

However, as Task~3.1 requires to also set the definitions of variables, unlike the initialization from Task~1, Task~3.1 transforms expressions into more complex objects that inherit from an interface to construct the Expr elements and set the \textit{def} and \textit{use} links accordingly. This interface is presented in Figure \ref{fig:IExpressionDFInfo}.

By using such mutable objects, dependencies may also specify how they are persisted in the output of the transformation rule via persistors. 

The \textit{MarkInstantiatingFor} method marks the assignment transformation rule as instantiating for the \verb|Expression2Text|-rule, i.e. whenever an Expression should be transformed to text and the Expression is an AssignmentExpression, \verb|AssignmentExpression2Text| is called instead to create the output of the \verb|Expression2Text|-rule. \verb|Expression2Text| is still called, but here it is empty and serves as a hub, only. Thus, the transformation of an expressions to text does not need to know which concrete expressions exist. The \textit{Require} methods specify dependencies to other model elements.

\subsection{Task 2: Control Flow Graph}
\label{ch:Solution:Task2}

\begin{figure}[ht]
	\centering
		\includegraphics[width=1.00\textwidth]{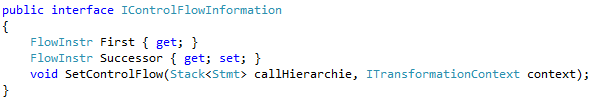}
	\caption{The interface of what is interesting regarding control flow}
	\label{fig:icfinfo}
\end{figure}

To derive the control flow graph, semantical information such as the first flow instruction within a statement has to be added to existing model elements. But it is not only data, but also the behaviour of how to set the control flow, that is important for this transformation. Thus, we define the necessary operations on statements that are necessary to derive the control flow, see Figure \ref{fig:icfinfo}.

At first, we need the first flow instruction since the entrance in a statement (\textit{First}). However, in some cases like empty blocks, this first flow instruction is not part of the current statement. Therefore, we need to tell what the next flow instruction after the current statement is (\textit{Successor}). Finally, the procedure of how to set the control flow for a statement also depends on a statement. A simple statement only sets the CfNext reference to the first inner flow instruction of the successor statement, whereas a jump statement sets a CfNext reference to the target jump label. However, some statements like \verb|break| and \verb|continue| cannot set their control flow successor without context. I.e., the successor of a continue statement is the test expression of the innermost loop that the continue is contained in. Additionally, the method also has the transformation context as parameter for tracing purposes.

\textsc{NMF Transformations} does not draw a difference between objects that are part of the model and objects that are just helpers. Thus, we can just use transformation rule that return implementations of the above interface for any statement. In an in-place transformation rule we only need to execute the resulting \textit{SetControlFlow} method for the \textsl{Method} element.

\subsection{Task 3.2: Deriving Data Flow}
\label{ch:Solution:Task3:Subtask3.2}

Deriving the data flow has been implemented in general purpose code. As \textsc{NMF Transformations} operates on POCOs, integrating this general purpose code in the transformation is just as easy as calling the algorithm. There is no conversion that has to be done. The algorithm works in that way that for a given variable, it follows the control flow everywhere until it arrives at a flow element that defines the same variable and thus the definition loses scope.

\subsection{Task 4: Validation}
\label{ch:Solution:Task4}

As the proposed query language is very simple, it suffices to solve the problem with regular expressions. To parse a command, we use the following regular expression:
\begin{verbatim}
(?<command>(cfNext|cfPrev|dfNext)):\s*
"(?<source>[^"]*)"\s*-->\s*"(?<target>[^"]*)"(;)?
\end{verbatim}
The validation application now just reads a target model and creates an internal hashtable with all the instructions that are contained in the model. Any validation string is then parsed with the above pattern and the application simply checks whether the asserted condition holds.

\section{Validation}
\label{ch:validation}

So far, the results have only been analyzed by reviewing the results for the output XMI files. All of the output XMI files validated successfully in Eclipse. 

The execution times on SHARE transforming the simple input models were really fast, see Table \ref{tab:ExecutionTimesOfThePerformanceTestCases}. The execution times for the smaller models were measured using hardware query performance counters for more exact time measurement. Otherwise execution times like these would be unable to measure.

The performance results show that the simplified algorithm to derive the data flow maybe is not that fast and possibly needs improvement. However, as this sort of algorithm was written in general purpose code, further optimization is not in the scope of this paper.


\section{Conclusion}
\label{ch:conclusion}

In this paper we have presented a solution to the TTC 2013 \textit{Flowgraphs} case based on \textsc{NMF Transformations}. It was not possible to support every bit of the transformation with \textsc{NMF}, as no support for iterative procedure is offered. However, it was easy to integrate general purpose code and cooperate with it. 

We suggest the high points of our solution as 
\begin{itemize}
	\item {\textbf{Good maintainability} through small change impacts as new metamodel elements are introduced}
	\item {\textbf{Excellent execution speed} with hardly measurable execution times except for the biggest models}
	\item {\textbf{Easy integration of general purpose code} whenever a task cannot be supported by \textsc{NMF} directly and has to be accomplished with general purpose code}
	\item {\textbf{Great tool support}, as \textsc{NMF Transformations} can reuse for example debugging, profiling, refactoring, testing and continuous integration support for C\#.}
\end{itemize}

The extension of the transformation from the initial solution to the updated solution supporting also unary expressions also showed that transformations in \textsc{NMF Transformations} are easy to maintain, as no existing code had to be changed to support this new requirement.

\phantomsection
\addcontentsline{toc}{chapter}{\bibname}

\iflanguage{english}{\bibliographystyle{eptcs}}	
{\bibliographystyle{eptcs}}	
												  

\bibliography{ttc2013_flowgraphs}



\appendix

\section{Appendix}

\begin{table}[H]
	\centering
		\begin{tabular}{|l|c|c|c|c|c|}
		\hline
		Test case & Reading input & Transformation & Derive Control Flow & Derive Data Flow & Writing output \\
		\hline
		\hline
0&	48.90ms &	4.48ms &	0.29ms &	0.08ms &	9.63ms \\
		\hline
1&	38.62ms &	4.45ms &	1.50ms &	0.09ms &	4.76ms \\
		\hline
2&	34.65ms &	4.35ms &	0.72ms &	0.11ms &	5.24ms \\
		\hline
3&	46.03ms &	3.01ms &	0.48ms &	0.05ms &	4.87ms \\
		\hline
4&	36.32ms &	2.72ms &	0.47ms &	0.04ms &	4.63ms \\
		\hline
5&	31.33ms &	3.71ms &	1.05ms &	0.05ms &	4.13ms \\
		\hline
6&	32.77ms &	4.60ms &	0.56ms &	0.04ms &	4.21ms \\
		\hline
7&	168.36ms &	27.15ms &	5.54ms &	5.73ms &	26.79ms \\
		\hline
8&	532.42ms &	97.08ms &	30.16ms &	68.26ms &	97.02ms \\
		\hline
9&	3,832.02ms &	1,103.49ms &	350.14ms &	6,345.40ms &	646.39ms \\
		\hline
10&	33.30ms &	1.94ms &	0.41ms &	0.03ms &	4.20ms \\
		\hline
11&	29.33ms &	3.36ms &	0.34ms &	0.05ms &	3.90ms \\
		\hline
		\end{tabular}
	\caption{Execution times of the test cases}
	\label{tab:ExecutionTimesOfThePerformanceTestCases}
\end{table}

\end{document}